\documentclass[twocolumn,showpacs,preprintnumbers,amsmath,amssymb,epsfig,floatfix,prx]{revtex4-1}

\usepackage{pdfpages}
\usepackage{pgffor}
\usepackage{etoolbox}
\makeatletter
\patchcmd{\@outputpage@head}{\@ifx{\LS@rot\@undefined}{}{\LS@rot}}{}{}{}
\makeatother

\usepackage{amsmath}
\usepackage{mathtools}
\usepackage{physics}
\DeclareMathAlphabet{\altmathcal}{OMS}{cmsy}{m}{n}
\usepackage{siunitx}
\usepackage{multirow}
\usepackage{amsfonts}
\usepackage{amssymb}
\usepackage{calligra}
\usepackage{calrsfs}
\DeclareMathAlphabet{\mathcalligra}{T1}{calligra}{l}{m}
\usepackage{epsfig}
\usepackage{graphicx}
\usepackage{dcolumn}
\usepackage{color}
\usepackage{natbib}  
\usepackage{hyperref}
\usepackage{breakurl}
\hypersetup{colorlinks=true, citecolor=blue, urlcolor=blue, linkcolor=blue}
\usepackage{bm}
\usepackage{booktabs}
\newcolumntype{C}{>{$}c<{$}}
\AtBeginDocument{
\heavyrulewidth=.08em
\lightrulewidth=.05em
\cmidrulewidth=.03em
\belowrulesep=.65ex
\belowbottomsep=0pt
\aboverulesep=.4ex
\abovetopsep=0pt
\cmidrulesep=\doublerulesep
\cmidrulekern=.5em
\defaultaddspace=.5em
}

\newcolumntype{L}[1]{>{\raggedright\arraybackslash}p{#1}}
\newcolumntype{C}[1]{>{\centering\arraybackslash}p{#1}}
\newcolumntype{R}[1]{>{\raggedleft\arraybackslash}p{#1}}
\newcommand{\C}{{\mathchoice{}{}{\scriptscriptstyle}{}C}}

\begin{document}
\title{Electrically tunable room-temperature ferromagnetism in CrBr$_3$}
\author{Chandan K. Singh}
\author{Mukul Kabir}
\email{mukul.kabir@iiserpune.ac.in}
\affiliation{Department of Physics, Indian Institute of Science Education and Research, Pune 411008, India}
\date{\today}

\begin{abstract}
The recent discovery of magnetic ordering in two-dimension has lead to colossal efforts to find atomically thin materials that order at high temperatures.  However, due to fundamental spin fluctuation in reduced dimension, the room-temperature ferromagnetism remains elusive. Here, we report a dramatic manipulation of magnetic ordering up to room temperature in the monolayer CrBr$_3$, within the first-principles Heisenberg XXZ model. The exchange and anisotropic magnetic interactions are externally modulated by a gate-induced charge carrier doping that triggers a nontrivial phase diagram. High-temperature ferromagnetism is associated with a substantial increase in both effective ferromagnetic exchange and overall magnetic anisotropy under experimentally attainable hole doping. In contrast, electron doping quickly switches the magnetic easy axis. The gate-tuneable room temperature ferromagnetism in CrBr$_3$ presents new possibilities in electrically controlled spintronic and magnetoelectric devices based on atomically thin crystals.
\end{abstract}
\maketitle


In contrast to the three-dimensional materials, magnetic anisotropy is indispensable for a finite-temperature phase transition in the two-dimension. 
The continuous rotational symmetry of the spin is the fundamental assumption to the Hohenberg-Mermin-Wagner theorem,~\citep{PhysRev.158.383,PhysRevLett.17.1133} which is broken by the magnetic anisotropy originated from the relativistic spin-orbit coupling in natural materials. 
With a finite anisotropy, the long-wavelength spin excitation becomes gapped, and a long-range magnetic ordering emerges at a non-zero temperature.~\citep{nature22060,nature22391,Gibertini2019} Moreover, the magnitude of the spin gap is dictated by the various anisotropic interactions, which determines the ordering temperature.~\citep{Lado_2017} Owing to the small spin-orbit coupling strength, the on-site single-ion anisotropy in real materials is much smaller, and the system cannot be analyzed by the Ising or XY models. Instead, an anisotropic Heisenberg model is a suitable descriptor that accounts for the spin fluctuations and results in significant renormalization of the ordering temperature. Thus, the two-dimensional magnetic materials operate at cryogenic temperatures,~\citep{nature22060,nature22391,Gibertini2019,Lado_2017,Kim11131,acs.nanolett.9b00553,s41565-019-0565-0,s41567-019-0651-0, adma.201808074,s41565-018-0063-9,s41586-018-0626-9} hindering potential utilization in classical and quantum information technologies. 

Since the intrinsic ferromagnetism has been experimentally witnessed in the two-dimensional van der Walls crystals of CrGeTe$_3$~\citep{nature22060} and CrI$_3$,~\citep{nature22391} a few other ferromagnetic materials have been identified. However, the Curie temperature in all these materials persisted much below the room-temperature.~\citep{Kim11131,acs.nanolett.9b00553,s41565-019-0565-0,s41567-019-0651-0, adma.201808074,s41565-018-0063-9,s41586-018-0626-9} For example, the insulating monolayers of chromium trihalides ferromagnetically order below the liquid-nitrogen temperature, with a maximum of 45 K in CrI$_3$.~\citep{nature22391,Kim11131,acs.nanolett.9b00553,s41565-019-0565-0} The Curie temperature of the bilayer CrGeTe$_3$ and monolayer Fe$_3$GeTe$_2$ are also much below the ambient temperature, at 30 and 130 K, respectively.~\citep{nature22060,s41563-018-0149-7} While manganese selenides are antiferromagnetic in bulk, the room-temperature ferromagnetism is observed in single-layer MnSe$_2$.~\citep{acs.nanolett.8b00683} However, it is not clear if the ferromagnetism is intrinsic or results from the defects and interfacial effects in MnSe$_2$ grown by molecular beam epitaxy on vdW substrates. Despite considerable attention, the high temperature ferromagnetism in $1T$-VSe$_2$ remains doubtful.~\citep{s41565-018-0063-9,acs.nanolett.8b01764,acs.jpcc.9b04281} Contradictory results of room-temperature ferromagnetism and high-temperature charge density wave phase that suppresses any magnetism have been reported. Therefore, the room-temperature ferromagnetism remains elusive, and the fundamental question of Curie temperature manipulation arises naturally, which may be achieved by externally modulating the exchange and anisotropic magnetic interactions.

Intrinsic magnetism may be manipulated by coupling to external perturbations such as strain,~\citep{PhysRevB.98.144411} nanoscale patterning,~\citep{acs.nanolett.8b02806} gating,~\citep{s41565-018-0121-3,s41565-018-0135-x,s41586-018-0626-9,PhysRevB.103.214411} chemical doping~\citep{acs.nanolett.9b03316} and intense light.~\citep{PhysRevLett.125.267205} Moreover, electric control of magnetism has colossal importance in nanoscale magnetic devices and has already proven effective in vdW materials. For example, the nature and strength of exchange interactions in Fe$_3$GeTe$_2$ and CrI$_3$ have been modified by injecting carriers through electrostatic gating.~\citep{s41565-018-0121-3,s41565-018-0135-x,s41586-018-0626-9,PhysRevB.103.214411} 
In insulating CrI$_3$, the ordering temperature, saturation magnetization, coercivity, and interlayer magnetic ordering have been manipulated by tuning the gate voltage.~\citep{s41565-018-0121-3,s41565-018-0135-x,PhysRevB.103.214411} The Curie temperature of the itinerant Fe$_3$GeTe$_2$ flakes shows a three-fold increase and beyond the room-temperature.~\citep{s41586-018-0626-9} Interaction with the femtosecond laser can also push the ordering temperature to a higher value, as indicated recently in a few-layer Fe$_3$GeTe$_2$.~\citep{PhysRevLett.125.267205} While some experimental success has been reported to manipulate the intrinsic magnetism, the microscopic mechanism down to the various magnetic interactions remains unexplained.

Charge doping of Mott insulators away from the half-filling may nontrivially modify the underlying magnetism. Indeed, here we report room-temperature ferromagnetism in hole-doped CrBr$_3$ monolayer. Predictions are made within a long-range anisotropic XXZ Heisenberg model, where the parameters are obtained from the first-principles calculations.  While monolayer CrBr$_3$ orders at about 30 K,~\citep{Kim11131,acs.nanolett.9b00553,PhysRevLett.124.197401,s41928-019-0302-6} experimentally feasible hole doping is predicted to drive the Curie temperature beyond 300 K. In contrast, electron doping has a negligible effect on the ordering temperature but easily flips the magnetic easy axis. We elaborate on the microscopic origin of the electrically controllable magnetic phase diagram and disentangle the effects of induced strain on carrier injection. The present results indicate that CrBr$_3$ warrants further experimental attention to test the current predictions.  

We consider the anisotropic XXZ Heisenberg model Hamiltonian with interactions beyond the first neighbour,~\citep{PhysRevB.103.214411} 
\begin{equation}
\altmathcal{H} = -\frac{1}{2} \sum_{k=1}^3 \sum_{\langle ij \rangle_k} \bigl(\altmathcal{J}_k \mathbf{S}_{i} \cdot \mathbf{S}_{j}  + \Lambda_k S_{i}^zS_{j}^z \bigr) - \sum_i \altmathcal{A}_z S_i^zS_i^z,
\nonumber
\end{equation}
where $\langle ij \rangle_k$ with $k=1, 2, 3$ represents the first, second and third neighbours in the honeycomb magnetic Cr$^{3+}$ ($S$ = 3/2) lattice. The second and third neighbour interactions are critical to characterize magnetism in two-dimensional magnets.~\citep{PhysRevB.103.214411} $\altmathcal{J}_k$ and $\Lambda_k$ are the corresponding isotropic and anisotropic exchange interactions. The neighbouring spins $\mathbf{S}_{i}$ and $\mathbf{S}_{j}$ are coupled ferromagnetically $\altmathcal{J}_k > 0$ or antiferromagnetically $\altmathcal{J}_k > 0$.  The spin-orbit coupling at anion produces symmetric anisotropic superexchange $\Lambda_k$, and fundamentally crucial for ordering. $\altmathcal{A}_z$ represents the on-site single-ion anisotropy, and $\altmathcal{A}_z > 0$ ($\altmathcal{A}_z < 0$) easy-axis (easy-plane) magnetism. We demonstrated earlier that such a spin model provides a quantitative description of the experimental results.~\citep{PhysRevB.103.214411}   

The parameters of the Heisenberg spin model are extracted from the relativistic first-principles calculations. Wave functions in the density functional theory are expressed within the projector augmented formalism~\citep{PhysRevB.50.17953} as implemented in the Vienna {\em ab initio} simulation package.~\citep{PhysRevB.47.558,PhysRevB.54.11169}. A plane-wave basis of 600 eV for the kinetic energy cut-off is used, and the spin-orbit coupling is considered. Local density approximation is used to represent the exchange-correlation energy, which is supplemented with the Hubbard-like on-site Coulomb interaction $U_{\rm Cr}$ (1 eV) for the localized 3$d$-electrons.~\citep{PhysRevB.57.1505} To describe all possible magnetic structures, we consider a 2$\times$2$\times$1 supercell, and the first Brillouin zone is sampled with a $\Gamma$-centred 17$\times$11$\times$1 Monkhorst-Pack $k$-grid.~\citep{PhysRevB.13.5188} Vacuum space of 20 \AA\ is assumed to minimize the interactions between the periodic images. The structures are optimized until all the force components are below 5$\times$10$^{-3}$ eV/\AA\ threshold.  The magnetic phase transition is studied using the Heisenberg Monte Carlo method on a periodic 50 $\times$ 50 lattice consisting of $10^4$ spins that minimize the finite-size effects. A random direction to a random three-dimensional spin $\mathbf{S}_i \rightarrow \mathbf{S}_i^{\prime}$ is considered using the Marsaglia procedure,~\citep{marsaglia1972} followed by an update within the Metropolis algorithm.~\citep{10.1063/1.1699114} To ensure thermal equilibrium, 2 $\times$ 10$^8$ Monte Carlo steps are employed at every temperature, and 192 independent simulations are used to reduce the statistical fluctuation. The Curie temperature $T_{\C}$ is calculated by fitting the magnetic order parameter, $m(T) = m_0(1-T/T_{\C})^{\beta}$, where $\beta$ is the critical exponent.

The magnetic Cr-atoms are arranged in a honeycomb lattice, surrounded by edge-sharing Br-octahedra. The calculated in-plane lattice parameter of 6.25 \AA\ is consistent with the single-layer and bulk CrBr$_3$ crystals.~\citep{science.aav1937,cryst7050121} Electronically CrBr$_3$ is a Mott insulator, and the half-filled $t_{2g}^3$ states are separated by 1.68 eV from the empty $e_{g}^0$ states, consistent with experimental data.~\citep{acs.nanolett.9b00553} Similar to CrI$_3$, the monolayer CrBr$_3$ is a ferromagnetic (FM) insulator following the Anderson-Goodenough-Kanamori rules.~\citep{PhysRev.79.350,GOODENOUGH1958287,KANAMORI195987} Ligand mediated superexchange interactions between the half-filled Cr-$t_{2g}^3$ levels with Cr$-$Br$-$Cr angle of 94$^\circ$ trigger FM interaction. Magnetic moments are primarily localized at the Cr-sites (2.93 $\mu_B$), which is in accordance with the experimental data.~\citep{JPSJ.15.1664} The zigzag antiferromagnetic (AFM) solution is the first excited state, $\Delta_{\rm Z}^{\rm FM}$ = 5.5 meV/CrBr$_3$, while the N\'eel and stripe AFM are $\Delta_{\rm N}^{\rm FM}$ = 11.1 and $\Delta_{\rm S}^{\rm FM}$ = 11.9 meV/CrBr$_3$ higher in energy. Due to the lower spin-orbit coupling at the halide site in CrBr$_3$, the anisotropic interactions are much smaller than the monolayer CrI$_3$.  

Monte Carlo simulations using the Heisenberg Hamiltonian $\altmathcal{H}$ reveal an FM phase transition at 30.5 K ($\beta = 0.226$), which is consistent with several experimental measurements of 27$-$34 K, such as magnetic circular dichroism,~\citep{Kim11131} helicity resolved photoluminescence,~\citep{acs.nanolett.9b00553,PhysRevLett.124.197401} magneto-optical Kerr effect,~\citep{PhysRevLett.124.197401} and  micromagnetometry.~\citep{s41928-019-0302-6} In contrast, a relatively lower $T_{\rm C}$ is estimated recently for hBN encapsulated monolayer and bulk samples.~\citep{s41563-020-0706-8} The sluggish dependence of $T_{\rm C}$ on layer thickness with $T_{\rm C}^{\rm bulk} = 37$ K indicates the 2D nature of magnetism and weaker interlayer magnetic coupling.~\citep{s41928-019-0302-6,JPSJ.15.1664}  
Exchange interactions beyond the first-neighbour with significant $\altmathcal{J}_2/\altmathcal{J}_1 \sim 0.2$ and $\altmathcal{J}_3/\altmathcal{J}_2 \sim -0.8$ cannot be neglected to explain phase transition. While the spin model with the first-neighbour interactions underestimates the $T_{\rm C}$ (22.5 K, $\beta$ = 0.244), the consideration of the interactions till the second-neighbour overestimates it to 37 K ($\beta$ = 0.247). Further, magnetism in CrBr$_3$ resides far away from the Ising model ($\altmathcal{A}_z \rightarrow \infty $) description due to the very small $\altmathcal{A}_z$ of 0.08 meV. Therefore, owing to severe spin-fluctuations, the $T_{\rm C}$ is renormalized from $T_{\rm C}^{\rm Ising}$ of 102 K, which is erroniously much higher than the experimental and the current predictions within the Heisenberg XXZ model.                                                                                                                                                                                                                                                                                                                                                        

\begin{figure}[!t]
\begin{center}
{\includegraphics[width=0.45\textwidth, angle=0]{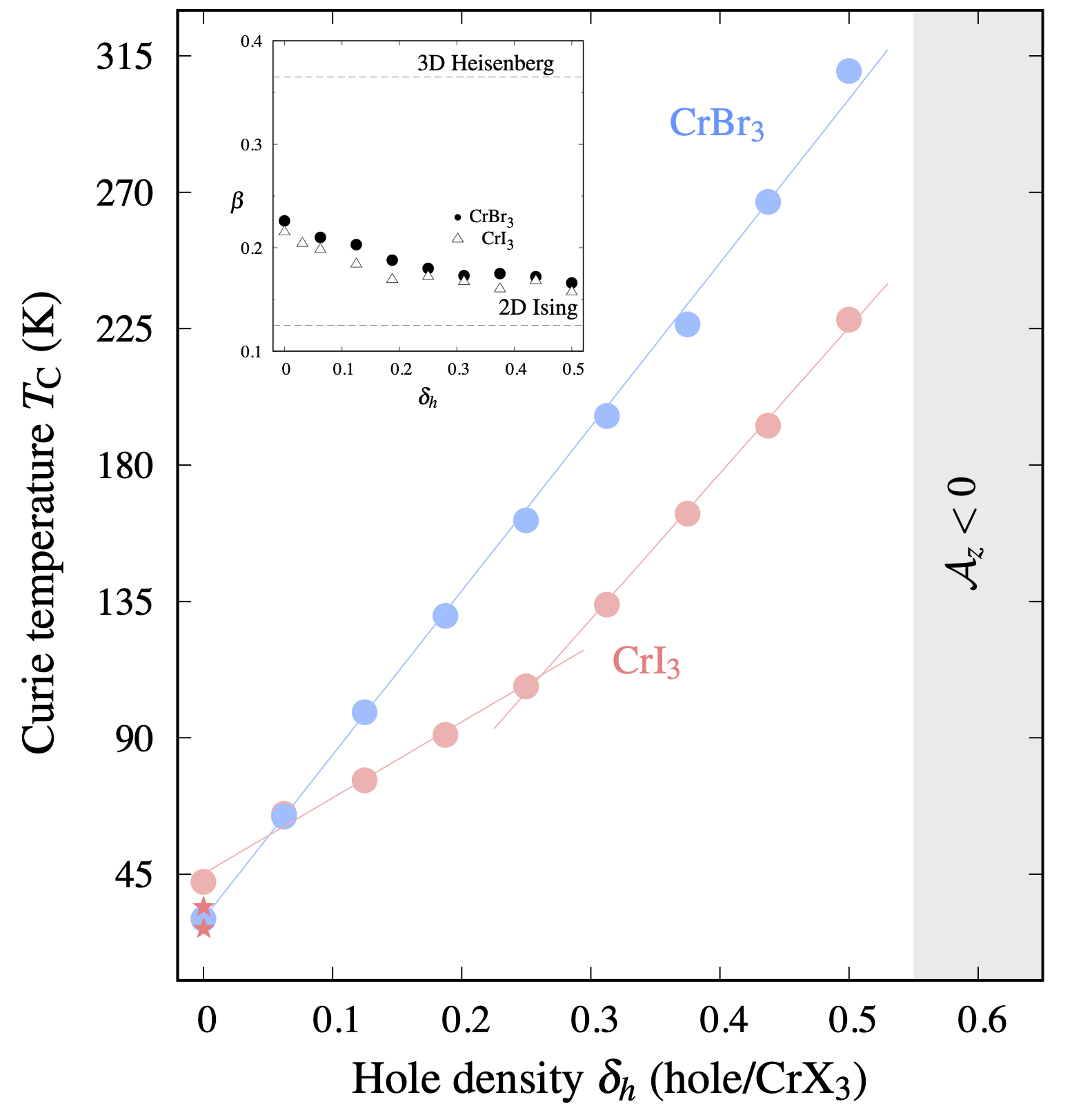}}
\caption{Electrical manipulation of magnetic ordering in monolayer Cr$X_3$. Calculated $T_{\rm C}$ is critically influenced by hole doping. The present results for the neutral monolayers are in quantitative agreement with the experimental $T_{\rm C}$, ~\citep{nature22391,s41928-019-0302-6,Kim11131,PhysRevLett.124.197401,acs.nanolett.9b00553} starts designate the experimental data for CrBr$_3$ monolayer. The evolution of $T_{\rm C}$ in CrBr$_3$ is compared with CrI$_3$. Calculated $T_{\C}$ linearly varies with hole density $\delta_h$, and room-temperature ferromagnetism is predicted in the monolayer  in CrBr$_3$  at $\delta_h = $ 0.5 hole/CrBr$_3$ $\sim$ 3.1 $\times 10^{14}$ cm$^{-2}$. Such high carrier density could be experimentally realised through ionic-gating or femtosecond laser pulse.~\citep{ncomms9826,nnano.2015.314,PhysRevLett.125.267205} The inset figure tracks the variation in magnetic critical exponent $\beta$ in monolayers, which approaches toward the 2D Ising value of 0.125 with increasing $\delta_h$. 
}
\label{fig:figure1}
\end{center}
\end{figure}

\begin{figure}[!t]
\begin{center}
{\includegraphics[width=0.48\textwidth, angle=0]{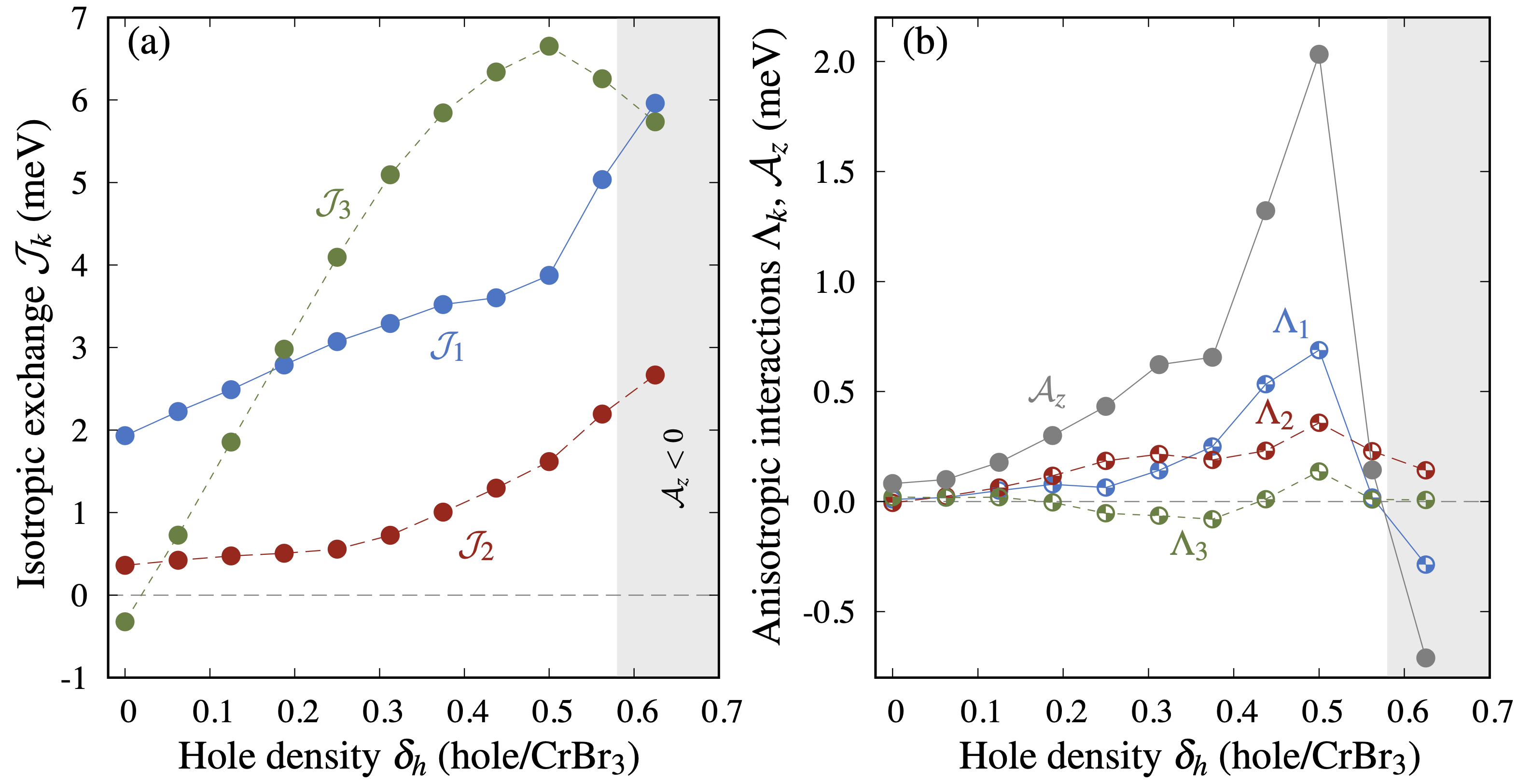}}
\caption{Charge doping alters the exchange $\altmathcal{J}_k$ and anisotropic interactions $\Lambda_k$ and $\altmathcal{A}_z$, which dictates the phase transition. (a) For the entire hole density $\delta_h$, all the isotropic exchanges remain ferromagnetic. Compared to $\altmathcal{J}_1$ and $\altmathcal{J}_2$, the variation of the third-neighbour $\altmathcal{J}_3$ is more substantial, which sharply increases with $\delta_h$, indicating the importance of the long-range interaction to describe magnetism. (b) Anisotropic interactions, especially the $\Lambda_1$, $\Lambda_2$ and $\altmathcal{A}_z$, show a strong dependence on $\delta_h$ and influences the ordering temperature.  However, at a very high hole density, $\delta_h > 0.6$ hole/CrBr$_3$ (shaded area), the Cr-atoms switches direction and prefers an in-plane orientation with $\altmathcal{A}_z < 0$. The corresponding magnetism becomes weak XY-type, and the corresponding FM $T_{\rm C}$ is expected to be renormalised.   
}
\label{fig:figure2}
\end{center}
\end{figure}

Electrical manipulation of magnetism in ferromagnets has drawn much attention due to its promise in spintronic devices with low power consumption.~\citep{Ohno35050040,science.1136629,nnano.2008.406,nnano.2015.22} The change in carrier density influences the exchange interactions and magnetic anisotropy. Further, the Heisenberg spin model is univocally applicable despite the insulator to metal transition under charge doping, as the localized Cr-moments describe the magnetism (Supplemental Material). We observe that the monolayer CrBr$_3$ orders above the room-temperature under sufficient hole doping (Figure~\ref{fig:figure1}). A remarkable ten-fold increase in $T_{\rm C} \sim $ 310 K is observed at an experimentally achievable hole density, $\delta_h \sim $ 0.5 hole/CrBr$_3$ $\sim$ 3.1 $\times 10^{14}$ cm$^{-2}$. Such high carrier density can be experimentally achieved in two-dimension via ionic-liquid gating or femtosecond laser pulse.~\citep{ncomms9826,nnano.2015.314,Dhoot11834,PhysRevLett.125.267205} CrBr$_3$ shows similar qualitative trend with CrI$_3$ but with a few quantitative differences (Figure~\ref{fig:figure1}). First, the maximum attainable $T_{\rm C}$ in CrBr$_3$ is much higher compared to 225 K in CrI$_3$ at a same hole density. Further, the doping-dependent $T_{\rm C}$ in CrI$_3$ is non-monotonous and could be divided into two-regimes (Figure~\ref{fig:figure1}). At low hole density, the dependance is comparatively much slower, which becomes sharper beyond $\delta_h > $ 0.25 hole/CrI$_3$. In contrast, for monolayer CrBr$_3$, calculated $T_{\rm C}$ varies linearly with $\delta_h$ within the entire doping range with $\frac{dT_{\rm C}}{d\delta_h} \sim$ 540 K/hole (Figure~\ref{fig:figure1}). The nature of magnetism changes at $\delta_h > $ 0.56 hole/CrBr$_3$, where the Cr-spins switches to the in-plane orientation, $\altmathcal{A}_z < 0$, and the corresponding $T_{\rm C}$ is expected to be suppressed in XY-spin system. The energy ordering of the excited AFM structures is also altered upon hole doping (Supplemental Material). Instead of zigzag AFM in the undoped case, the stripe AFM structure becomes the first excited state at $\delta_h > $ 0.1 hole/CrBr$_3$. The critical exponent $\beta$ for the undoped CrBr$_3$ is in between the corresponding values for the 3D Heisenberg and 2D Ising models (inset of Figure~\ref{fig:figure1}). With increasing $\delta_h$, $\beta$ decreases toward the 2D Ising value as the overall anisotropy becomes stronger (Figure~\ref{fig:figure2}). The CrI$_3$ monolayer also exhibits the same qualitative trend.  
Contrary to the hole doping, electron doping is not supportive in the present context of $T_{\rm C}$ manipulation. However, it renders a rich magnetic phase diagram. Multiple phase transitions,  FM ($\altmathcal{A}_z > 0$) $\rightarrow$ FM ($\altmathcal{A}_z < 0$) $\rightarrow$ stripe AFM ($\altmathcal{A}_z > 0$) are observed with increasing $\delta_e$ (Supplemental Material). Biaxial tensile and compressive strain is induced in the hexagonal lattice under electron and hole doping, respectively. However, the $T_{\rm C}$ exhibits a weak and nonmonotonous dependence on the strain, and the observed hole-dependent increase in ordering temperature is entirely of electronic origin (Supplemental Material).

The microscopic interpretation of carrier-induced enhancement in ordering temperature (Figure~\ref{fig:figure1}) is intriguing. Charge doping severely alters the exchange $\altmathcal{J}_k$ and anisotropic interactions $\Lambda_k$ and $\altmathcal{A}_z$ (Figure~\ref{fig:figure2}), which drives the magnetic phase transition. For the entire hole density $\delta_h$, all the isotropic exchanges remain ferromagnetic in CrBr$_3$ monolayer. Compared to $\altmathcal{J}_1$ and $\altmathcal{J}_2$, the variation of the third-neighbour $\altmathcal{J}_3$ is more substantial [Figure~\ref{fig:figure2}(a)], which sharply increases with $\delta_h$, indicating the importance of the long-range interaction to describe magnetism. Anisotropic interactions, especially the $\Lambda_1$, $\Lambda_2$ and on-site $\altmathcal{A}_z$, increases significantly with $\delta_h$ [Figure~\ref{fig:figure2}(b)] and influences the ordering temperature. Using gate-induced coercivity data in few-layer Fe$_3$GeTe$_2$ such carrier induced modulation in magnetic anisotropy has been discussed.~\citep{s41586-018-0626-9} At a very high hole density, $\delta_h > 0.6$ hole/CrBr$_3$, the Cr-atoms switches direction and prefers an in-plane orientation with $\altmathcal{A}_z < 0$. The corresponding magnetism becomes weak XY-type, and the corresponding FM $T_{\rm C}$ is expected to be renormalized.

\begin{figure}[!t]
\begin{center}
{\includegraphics[width=0.45\textwidth, angle=0]{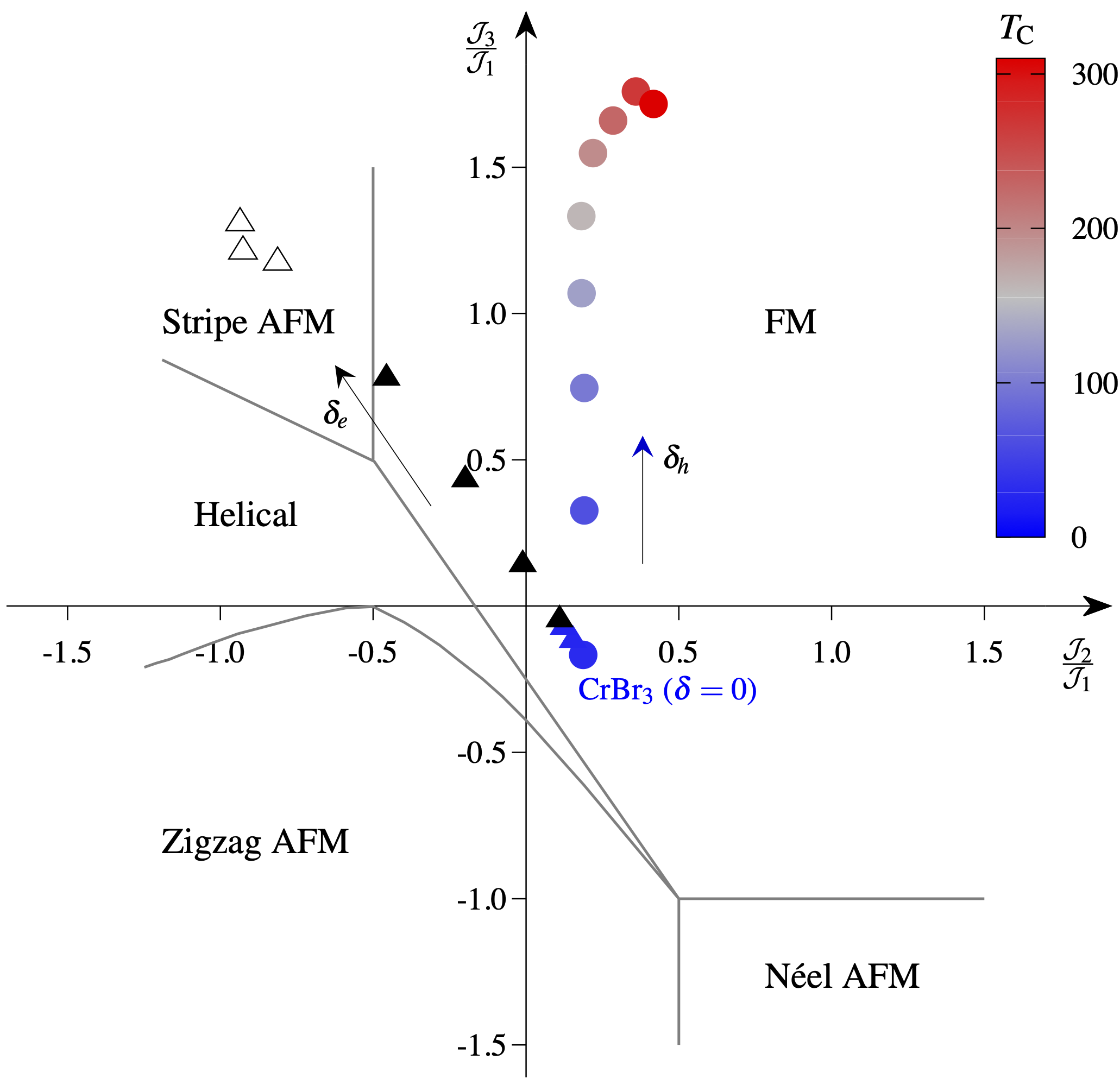}}
\caption{Calculated first-principles data, isotropic exchange interactions $\frac{\altmathcal{J}_2}{\altmathcal{J}_1}$, $\frac{\altmathcal{J}_3}{\altmathcal{J}_1}$ and the corresponding $T_{\rm C}$  are plotted on the honeycomb lattice $\altmathcal{J}_1$-$\altmathcal{J}_2$-$\altmathcal{J}_3$ XXZ  Heisenberg phase diagram with ferromagnetic first-neighbour interaction (adopted from the Ref.~\cite{s100510170273}). Regardless of the nature of carrier doping, $\altmathcal{J}_1$ is always FM for CrBr$_3$. Hole (circle) and electron (triangle) doping with increasing density are traced. The CrBr$_3$ monolayer remains FM under any hole density $\delta_h$, and the continuous increase in Curie temperature is traced with color palette. With increasing electron density $\delta_e$, multiple phase transitions,  FM ($\altmathcal{A}_z > 0$; blue $\blacktriangle$) $\rightarrow$ FM ($\altmathcal{A}_z < 0$; black $\blacktriangle$) $\rightarrow$ stripe AFM ($\altmathcal{A}_z > 0$; black $\triangle$) are observed. 
}
\label{fig:figure3}
\end{center}
\end{figure}

To gain further understanding, we project the present carrier-induced phase diagram onto the classical honeycomb $\altmathcal{J}_1$-$\altmathcal{J}_2$-$\altmathcal{J}_3$ XXZ Heisenberg model with FM $\altmathcal{J}_1$,~\citep{s100510170273,PhysRevB.97.134409} and find an excellent agreement (Figure~\ref{fig:figure3}). The undoped CrBr$_3$ ($\frac{\altmathcal{J}_2}{\altmathcal{J}_1} \sim$ 0.2 and $\frac{\altmathcal{J}_3}{\altmathcal{J}_1} \sim$ 0.17) resides near the narrow helical and vast zigzag AFM phases. While both $\altmathcal{J}_1$ and $\altmathcal{J}_2$ increase with $\delta_h$ [Figure~\ref{fig:figure2}(a)], the $\frac{\altmathcal{J}_2}{\altmathcal{J}_1}$ ratio almost remains unaltered (Figure~\ref{fig:figure3}). In contrast, the $\frac{\altmathcal{J}_3}{\altmathcal{J}_1}$ increases monotonically in the range $-0.2 < \frac{\altmathcal{J}_3}{\altmathcal{J}_1} < 1.75$. Therefore, consistent with the classical phase diagram, at any hole density, CrBr$_3$ remains FM, and the corresponding $T_{\rm C}$ increases with $\delta_h$ (Figure~\ref{fig:figure3}). The phase diagram upon electron doping is rather complex. While the first-neighbour $\altmathcal{J}_1$ decreases with electron density and remains FM (Supplemental Material), the ratio $\frac{\altmathcal{J}_2}{\altmathcal{J}_1}$ decreases with $\delta_e$ and becomes negative (Figure~\ref{fig:figure3}). In contrast, $\frac{\altmathcal{J}_3}{\altmathcal{J}_1}$ monotonically increases with $\delta_e$ in the range $-0.2 < \frac{\altmathcal{J}_3}{\altmathcal{J}_1} < 1.1$. Such combinatorial variations in $\frac{\altmathcal{J}_2}{\altmathcal{J}_1}$ and $\frac{\altmathcal{J}_3}{\altmathcal{J}_1}$ trigger a rich phase diagram. (i) CrBr$_3$ remains ferromagnetic at $\delta_e < $ 0.02 electron/CrBr$_3$, where the $T_{\rm C}$ has a weaker dependence and monotonically decreases to 25 K. (ii) At 0.02 $< \delta_e <$ 0.2 electron/CrBr$_3$, the monolayer remains FM, but the Cr-spins switches to the in-plane direction ($\altmathcal{A}_z < 0$). A weak XY-system represents the magnetism, and the corresponding FM $T_{\rm C}$ is anticipated to be renormalised.  Therefore, regardless the sign of $\altmathcal{A}_z$, an FM solution emerges for $-0.5 < \frac{\altmathcal{J}_2}{\altmathcal{J}_1} < -0.2$ and $1 < \frac{\altmathcal{J}_3}{\altmathcal{J}_1} > -0.2$. (iii) At higher density, $\delta_e > $ 0.2 electron/CrBr$_3$, another phase transition is observed when a stripe AFM with $\altmathcal{A}_z > 0$ becomes the ground state, which remains the same with further increase in $\delta_e$ with $\frac{\altmathcal{J}_2}{\altmathcal{J}_1} < -0.5$ and $\frac{\altmathcal{J}_3}{\altmathcal{J}_1} > 1$. 

Since the localized Cr-moments still describe the magnetism, the multiorbital Kugel-Khomskii mechanism of superexchange can corroborate the carrier induced ferromagnetism.~\citep{Kugel1982} In the undoped monolayer, the FM $t_{2g}^3-e_g^0$ interaction dominates over the AFM $t_{2g}^3- t_{2g}^3$. With increasing $\delta_h$, the partial hole $\delta_0$ in the $t_{2g}^{3-\delta_0}$ level destabilizes the AFM $t_{2g}- t_{2g}$ interaction and a stronger FM solution emerges.
Spin excitation becomes gapped due to anisotropic magnetic interactions, and thus, the spin-wave analysis provides further understanding into ordering. The Heisenberg XXZ Hamiltonian can be rewritten in Holstein-Primakoff bosons and recast the spin operators with the magnon annihilation and creation operators.~\citep{PhysRev.58.1098} At low-temperature ($k_BT \ll \altmathcal{J}_k$ and $S^z \simeq S$), we use linear approximation, $S^x + iS^y = \sqrt{2S}a$, $S^x - iS^y = \sqrt{2S}a^{\dagger}$, and $S^z = S - a^{\dagger}a$. The corresponding spin-wave Hamiltonian describe magnon excitation with energy, $\epsilon_0 = 2S\altmathcal{A}_z + \sum_k n_k \altmathcal{J}_kS + \sum_k n_k \Lambda_k S$, where $n_k$ is the number of first, second and third nearest-neighbours on the honeycomb lattice. This results in zero-temperature magnon gap at $\Gamma$ point, 
$\Delta_0 = 2S\altmathcal{A}_z + 3S(\Lambda_1 + 2\Lambda_2 + \Lambda_3)$, which yields  0.34 meV for the undoped CrBr$_3$. Owing to the increase in on-site and exchange anisotropies (Fig.~\ref{fig:figure2}), the gap $\Delta_0$ rapidly increases with $\delta_h$ (Supplemental Material) indicating magnetic ordering at elevated temperatures.

Electrically controlled magnetism can be utilized to design various energy-efficient spin-based devices. A spin field-effect transistor (FET)  can be realized,~\citep{10.1063/1.102730,s41928-019-0232-3,10.1063/5.0014865} where the gate voltage controls the direction of the magnetic easy axis while longitudinal in-plane source-drain voltage produces spin current. 
Upon electron doping $\delta_e > 1.2 \times 10^{13}$ cm$^{-2}$, the CrBr$_3$ monolayer remains FM but the easy axis switches from out-of-plane to in-plane direction. Reorientation of spin at the undoped-doped interface will significantly modulate the device resistance. The spin-FET is naturally in the on-state with low-resistance, while gate-controlled electron doping beyond a critical doping creates a magnetic ($\Uparrow | \Rightarrow | \Uparrow$) interface resulting in the high-resistance state due to strong scattering. Besides, the 2D CrBr$_3$ can also be employed in all-vdW magnetic tunnel junction (MTJ), where graphene or h-BN may be used as the barrier.~\citep{Song1214,acs.nanolett.8b01278} The critical advantage will be that the CrBr$_3$-MTJ may be voltage-controlled and operate at a high temperature.

In conclusion, we show that charge doping in Mott insulating CrBr$_3$ induces nontrivial evolution in magnetism with a rich phase diagram that includes room-temperature ferromagnetism. While the undoped CrBr$_3$ monolayer orders around 30 K, the Curie temperature monotonically increases above room temperature within experimentally achievable hole density. The microscopic origin of the phase diagram is explained in terms of the evolution of long-range exchange and anisotropic parameters obtained from the first-principles calculations. Voltage-controlled room temperature ferromagnetism in atomically thin CrBr$_3$ provides unprecedented opportunities in two-dimensional spintronics, magnetoelectrics, as well as in the context of emergent quantum states.  

M.K. acknowledges funding from the Indian Science and Engineering Research Board through EMR/2016/006458 grant. We gratefully acknowledge the support and resources provided by the PARAM Brahma Facility at the Indian Institute of Science Education and Research, Pune under the National Supercomputing Mission of Government of India.


%

\clearpage
\foreach \x in {1, 2, 3, 4}
{%
\clearpage


\section*{Supplementary material}

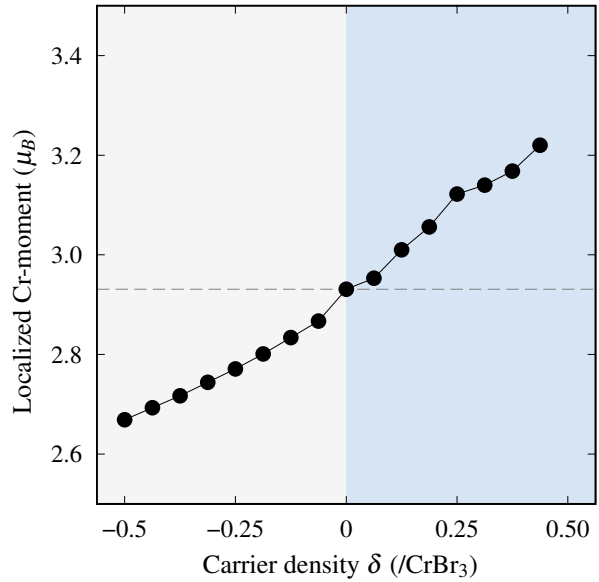

FIG. S1. The magnetic moment in the undoped CrBr$_3$ is localized at the Cr$^{3+}$-sites (2.93 $\mu_B$) in agreement with the experimental results. The scenario remains the same under charge doping, and the localized Cr-moments still describe the magnetism. Electron and hole doping only slightly perturbs the Cr-moment.

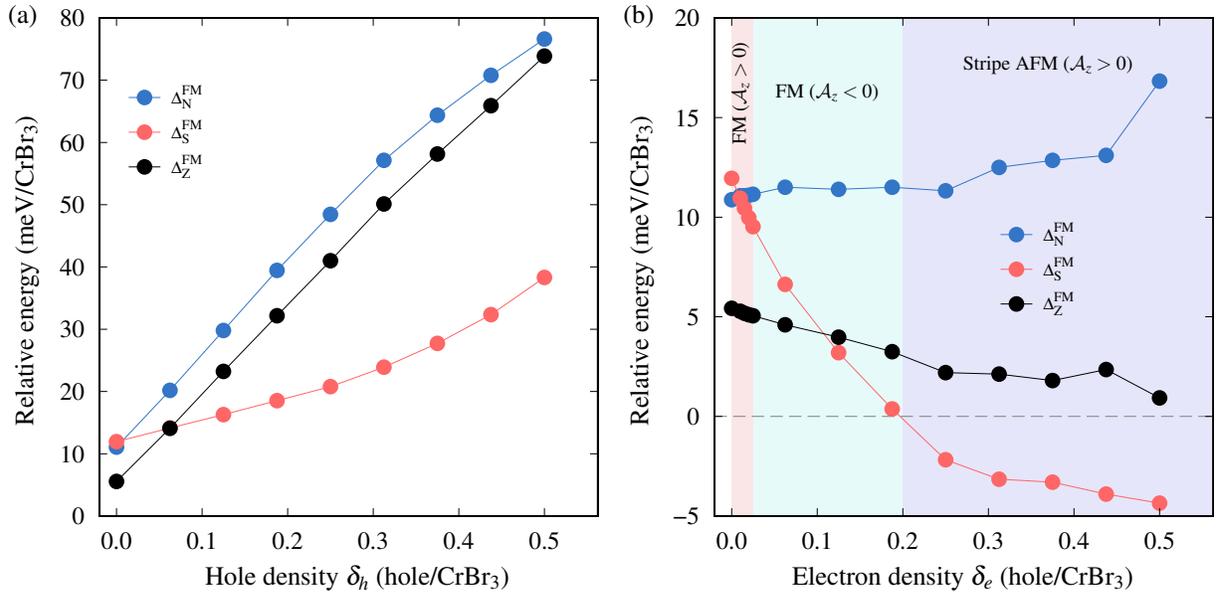

FIG. S2. Relative energies of various antiferromagnetic configurations with respect to the ferromagnetic state, for (a) hole and (b) electron doping. With increasing $\delta_h$, stipe and zigzag AFM switch energy ordering, and stripe AFM becomes the first excited state for $\delta > 0.1$ hole. Electron doped side of the phase diagram is rich, and the FM state with out-of-plane spin orientation ($\mathcal{A}_z > 0$) exists only with a narrow region of electron doping. With further increase in $\delta_e$ density, FM ($\mathcal{A}_z > 0$) → FM ($\mathcal{A}_z < 0$) → stripe AFM FM ($\mathcal{A}_z > 0$) is observed.

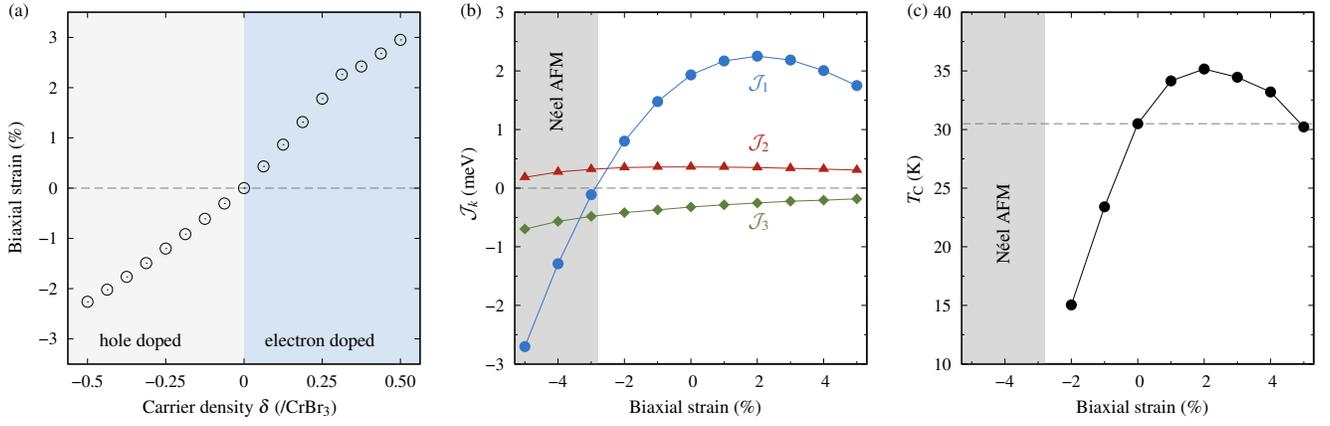

FIG. S3. (a) Biaxial tensile and compressive strain is induced in monolayer CrBr$_3$ due to electron and hole doping, respectively. (b) While the nature of $\mathcal{J}_2$ (FM) and $\mathcal{J}_3$ (AFM) remains unaltered, the variation in $\mathcal{J}_1$ is significant under strain. Further, the $\mathcal{J}_1$ becomes AFM at a high compressive strain beyond 3%, and a Néel AFM state emerges. (c) FM ordering temperature $T_\mathrm{C}$ show a relatively weak and nonmonotonous dependence on the induced strain.

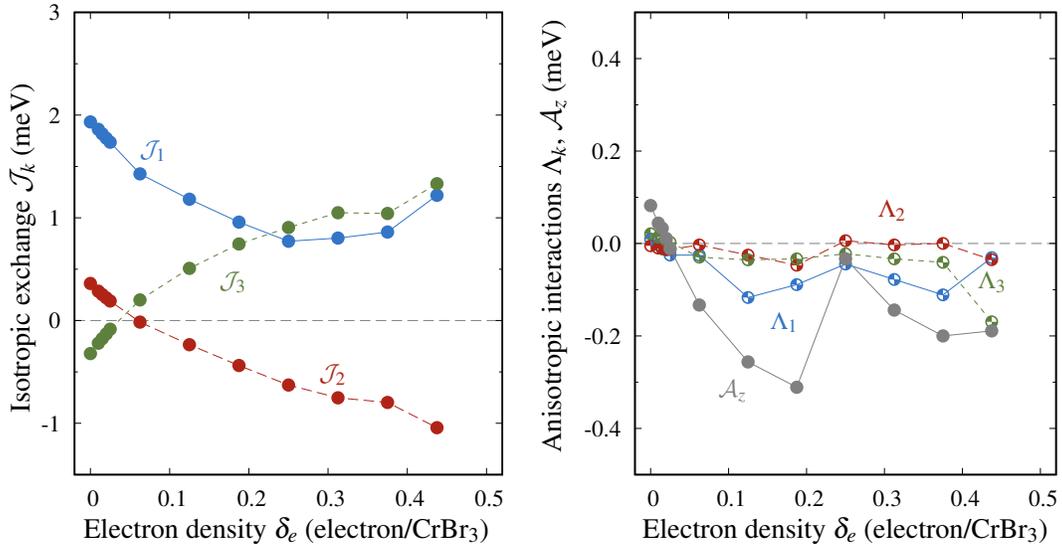

FIG. S4. The dependence of isotropic and anisotropic magnetic interaction upon electron doping is shown. The nature of FM $\mathcal{J}_2$ and AFM $\mathcal{J}_3$ switches beyond a small electron density of about 0.05 electron/CrBr$_3$. Most notably, the easy axis quickly switches from out-of-plane to in-plane upon electron doping.

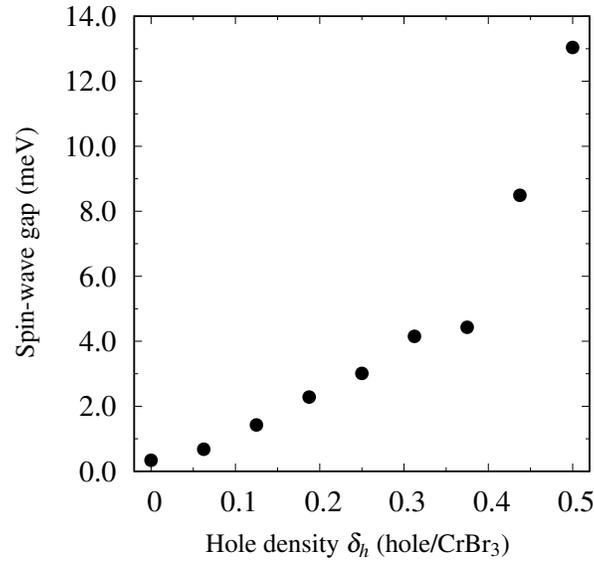

FIG. S5. The spin excitation gap, which is responsible for the finite-temperature phase transition in the two-dimension, is calculated within the linear spin-wave formalism of Holstein-Primakoff. The calculated spin-wave gap rapidly increases with hole density $\delta_h$, indicating a concurrent rise in the Curie temperature.

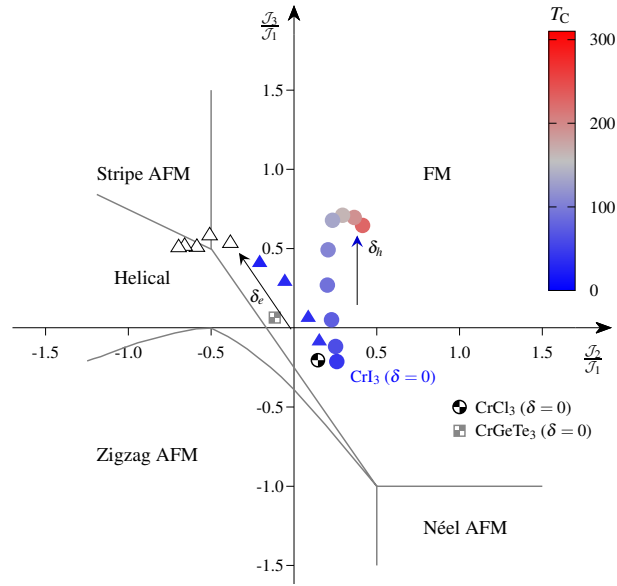

FIG. S6. The CrI$_3$ phase diagram is qualitatively similar to the CrBr$_3$ described in the main text. Similar to the CrBr3, the undoped monolayers of CrI$_3$, CrCl$_3$, and CrGeTe$_2$ lives near the helical and zigzag AFM phase boundary.


}

\end{document}